\documentclass[twocolumn]{ceurart}

\usepackage{float}
\usepackage{amsmath, amsfonts}
\usepackage{xspace}

\usepackage{orcidlink}

\usepackage{graphicx}
\usepackage{subcaption}

\usepackage[inline]{enumitem}
\setlist*[itemize]{labelindent=10pt, itemindent=0pt, leftmargin=*}

\usepackage{booktabs}
\usepackage{multirow}
\usepackage{tabularx}

\usepackage{wasysym}

\usepackage{pgfplots}
\pgfplotsset{compat=1.18}
\usepgfplotslibrary{statistics}

\usepackage{listings}
\usepackage{iftex}

\ifPDFTeX
  \usepackage[T1]{fontenc}
  \usepackage[utf8]{inputenc}
  \usepackage{zi4} 
\else
  \usepackage{fontspec}
\fi

\definecolor{vscode-bg}{HTML}{FFFFFF}
\definecolor{vscode-fg}{HTML}{1F2328}

\definecolor{vscode-keyword}{HTML}{0000FF}
\definecolor{vscode-string}{HTML}{A31515}
\definecolor{vscode-comment}{HTML}{008000}
\definecolor{vscode-number}{HTML}{098658}
\definecolor{vscode-func}{HTML}{795E26}
\definecolor{vscode-type}{HTML}{267F99}

\definecolor{LightGrey}{HTML}{BFBFBF}

\lstdefinelanguage{CPSLint}{%
      morekeywords={import, from, export, to, csv, is, skip, impute, in, real, nat, int, uart, str, bool, critical, sorted, unique, last, next, mean, median, interpolation, linear, time, zero, quadratic, cubic, barycentric, piecewise_polynomial, pchip, akima, cubicspline, from_derivatives, regex, empty, out, of, order, cut, when, inspect, using, perform},
      morestring=[b]',
}

\lstdefinelanguage{Rascal}{%
      morekeywords={syntax, data, module, alias, list, tuple, lrel, str, real, bool},
      morestring=[b]",
}

\lstdefinelanguage{yaml}{%
      morekeywords={inputFolder, outputFolder, pythonCommand, runWith},
      morecomment=[l]{\#},
}

\lstdefinelanguage{PythonExtended}[]{Python}{
  morekeywords={as},
}

\usepackage[nameinlink]{cleveref}

\usepackage{csquotes}
\usepackage{textcomp}

\usepackage{balance}

\newcommand{\CPSLint}{\textsf{CPSLint}\xspace}

\copyrightyear{2026}
\copyrightclause{Copyright for this paper by its authors. Use permitted under Creative Commons License Attribution 4.0 International (CC BY 4.0).}

\conference{Post-proceedings of BENEVOL 2025}

\begin{document}
\sloppy

\title{Implementing CPSLint: A Data Validation and Sanitisation Tool for Industrial Cyber-Physical Systems}
\shorttitle{Implementing CPSLint}

\author[1]{Uraz Odyurt}[%
orcid=0000-0003-1094-0234,
email=u.odyurt@utwente.nl,
]

\author[2]{\"{O}mer Sayilir}[%
orcid=0009-0009-8860-2316,
email=o.f.sayilir@utwente.nl,
]

\author[2]{Mari\"{e}lle Stoelinga}[%
orcid=0000-0001-6793-8165,
email=m.i.a.stoelinga@utwente.nl,
]

\author[2]{Vadim Zaytsev}[%
orcid=0000-0001-7764-4224,
email=vadim@grammarware.net,
]

\address[1]{%
Dynamics Based Maintenance, ET Faculty, University of Twente, The Netherlands
}

\address[2]{%
Formal Methods and Tools, EEMCS Faculty, University of Twente, The Netherlands
}

\begin{abstract}
Raw datasets are often too large and unstructured to work with directly, and require a data preparation phase. The domain of industrial Cyber-Physical Systems (CPSs) is no exception, as raw data typically consists of large time-series data collections that log the system’s status at regular time intervals. The processing of such raw data is often carried out using ad hoc, case-specific, one-off Python scripts, often neglecting aspects of readability, reusability, and maintainability. In practice, this can cause professionals such as data scientists to write similar data preparation scripts for each case, requiring them to do much repetitive work. We introduce \emph{\CPSLint}, a Domain-Specific Language (DSL) designed to support the data preparation process for industrial CPS. \CPSLint raises the level of abstraction to the point where both data scientists and domain experts can perform the data preparation task. We leverage the fact that many raw data collections in the industrial CPS domain require similar actions to render them suitable for data-centric workflows. In our DSL one can express the data preparation process in just a few lines of code. \emph{\CPSLint} is a publicly available tool applicable for any case involving time-series data collections in need of sanitisation.

\end{abstract}


\maketitle


\section{Introduction}
\label{sec:introduction}
Data from Cyber-Physical Systems (CPS) is crucial for gaining insight into the functioning of these complex systems, both in the form of real-time fault detection and postmortem analysis of failures. Examples of CPS are high-tech production machinery, e.g., semiconductor photolithography machines, die bonder machines, and so forth. Runtime traces from these systems can often contain corrupted data originating from, for example, faulty sensor readings, missing readings due to sensor outages, or even human error while managing this data. When performing diagnostics with CPS data, it is important to sanitise the data and ensure such corruptions would not interfere with the intended data processing. In the context of industrial CPS, data sanitisation requires both domain knowledge and programming skill, thus often requiring both a domain expert and a data scientist to be involved in performing the task. Besides the fact that it requires multiple domains of expertise, the code written to sanitise a specific dataset is often not re-applicable to other cases, creating a situation in which time and effort are spent writing single-use Python scripts.

Ideally speaking, such data sanitisation would be designed best if it is done by a domain expert who possesses programming skills as well. As this is seldom the case, having a simplified and contained language instead of a general programming language, e.g., Python, would enable domain experts for this task. Domain-Specific Languages (DSLs) are a good fit when it comes to context-specific language design.

Following this rationale, we introduce \CPSLint, a DSL built to aid the data sanitisation process. The language raises the level of abstraction for performing data sanitisation, making it accessible to domain experts and allowing them to perform this task without requiring substantial programming knowledge. It also has a relatively compact declarative syntax, making it also a suitable tool for data scientists by significantly reducing the amount of code they would need to write to sanitise a dataset.

\begin{figure}[htbp]
	\centering
	\includegraphics[width=0.9\linewidth]{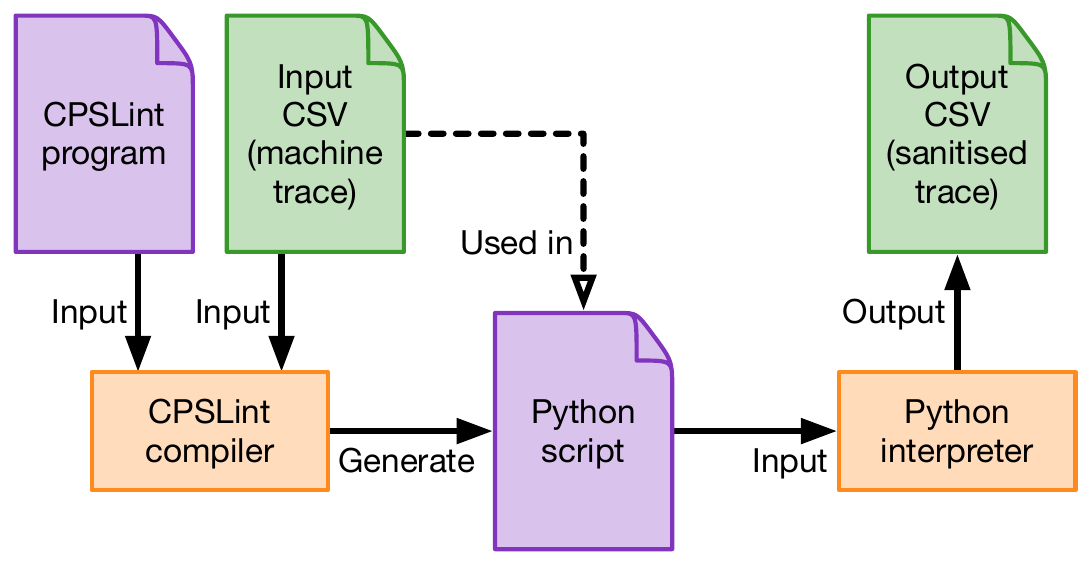}
	\caption{A typical \CPSLint workflow for sanitising industrial CPS machine traces.}
	\label{fig:pipeline_simple}
\end{figure}

\Cref{fig:pipeline_simple} depicts an example of a typical \CPSLint workflow. The compiler takes a \CPSLint specification and a raw CSV file as input. Our implementation then uses these to generate a human-readable Python script, capable of sanitising the raw CSV according to the provided definition. This script can then be run using a Python interpreter to obtain a sanitised CSV.

\paragraph*{Cyber-physical systems}
are complex systems combining cyber elements, i.e., computational and networked components, with interactions in the physical domain. Their design often integrates multitudes of subsystems, working in tandem to perform specific, specialised tasks. Thus, CPS are purpose-built machines. The industrial variants exhibit a significantly higher level of complexity and precision, excelling in high-yield production for high-tech industries. It is standard practice to constantly monitor industrial CPS for machine behaviour tracking and fault detection/prevention. Time-series data collection is the de facto monitoring data format for these devices.

\paragraph*{Domain-specific languages}
are software languages with a higher level of abstraction than General-purpose Programming Languages (GPLs)~\cite{Mernik:2005:WHDD}. They are designed for specific tasks and enable domain experts to perform programming using familiar terms and concepts. The output of DSLs need not be executable, as in the case of \LaTeX{}, YAML, and HTML, which are all markup languages. In short, DSLs make it easier for non-programmer domain experts to perform tasks programmatically within their domain than GPLs do.

\paragraph*{Contribution}
We showcase a DSL that was built based on similarities across the data validation and sanitisation domains, combined with information from the cyber-physical domain. The current implementation of \CPSLint provides a small but powerful declarative data sanitisation language. The compactness of our current language implementation is demonstrated by a relatively small codebase, i.e., 1978 lines of code. The data involved in our demonstrations, alongside the \CPSLint code, are openly available~\cite{Sayilir:2025:CODE}\footnote{Also available at: \url{https://github.com/omersayilir75/CPSLint}}.

The theoretical background leading to \CPSLint as a tool and more detailed information on use-cases, alongside systematic benchmarking of computational performance for different modes, is provided in the SLE 2026 paper~\cite{Odyurt:2026:DSLP}.

\section{Data}
\label{sec:data}
While it can be adapted in other scenarios, \CPSLint is primarily designed to work with industrial CPS data. The term `data' in this context loosely refers to the combination of sensor data, alongside contextual logs. Accordingly, sensor data is recorded as time-series, covering one or more metrics. While the recoding is commonly done at a high frequency, e.g., 1~kHz, each metric will be recorded with a frequency supported by the respective sensor and collection logic. As an example relevant to our demonstration use-case, power metrics such as voltage and current can be collected from different components on a system. Depending on the desired granularity or available support, such collections could happen at the component level, the controller level (for instance motor controller), at the subsystem level, or at the system level.

The contextual logs are expected to contain the signalling information related to a system's activity, i.e., start and stop signals per task or subtask. While being a common practice, the granularity at which such signals are recorded depends on the designer's foresight. It must be mentioned that given access to the underlying software element of an industrial CPS, it is relatively easy to extend contextual logging.

\subsection{Data format}
When considering time-series data, the de facto format is Comma-Separated Values (CSV) files. The \emph{timestamp} column is almost always present, with rare exceptions that consider an ordered index instead. One or more metric columns are included as well, depending on the modality of collected data. For instance, CSV files from our demonstration use-case include voltage, current and energy readings, as detailed below:
\begin{itemize}[nosep]
    \item Timestamps (S) column with values in ms
    \item Voltage (V) column with values recorded at 1~kHz
    \item Current (A) column with values recorded at 4~kHz
    \item Energy (J) column with values reflecting consumed energy 
    \item Universal Asynchronous Receiver-Transmitter (UART) messages as string data
\end{itemize}

The UART message column is populated sparsely, only indicating subtask starts and stops, or transmitting information such as processor core temperature.

\subsection{Data compartmentalisation}
\begin{figure}[thbp]
	\centering
	\includegraphics[width=\linewidth]{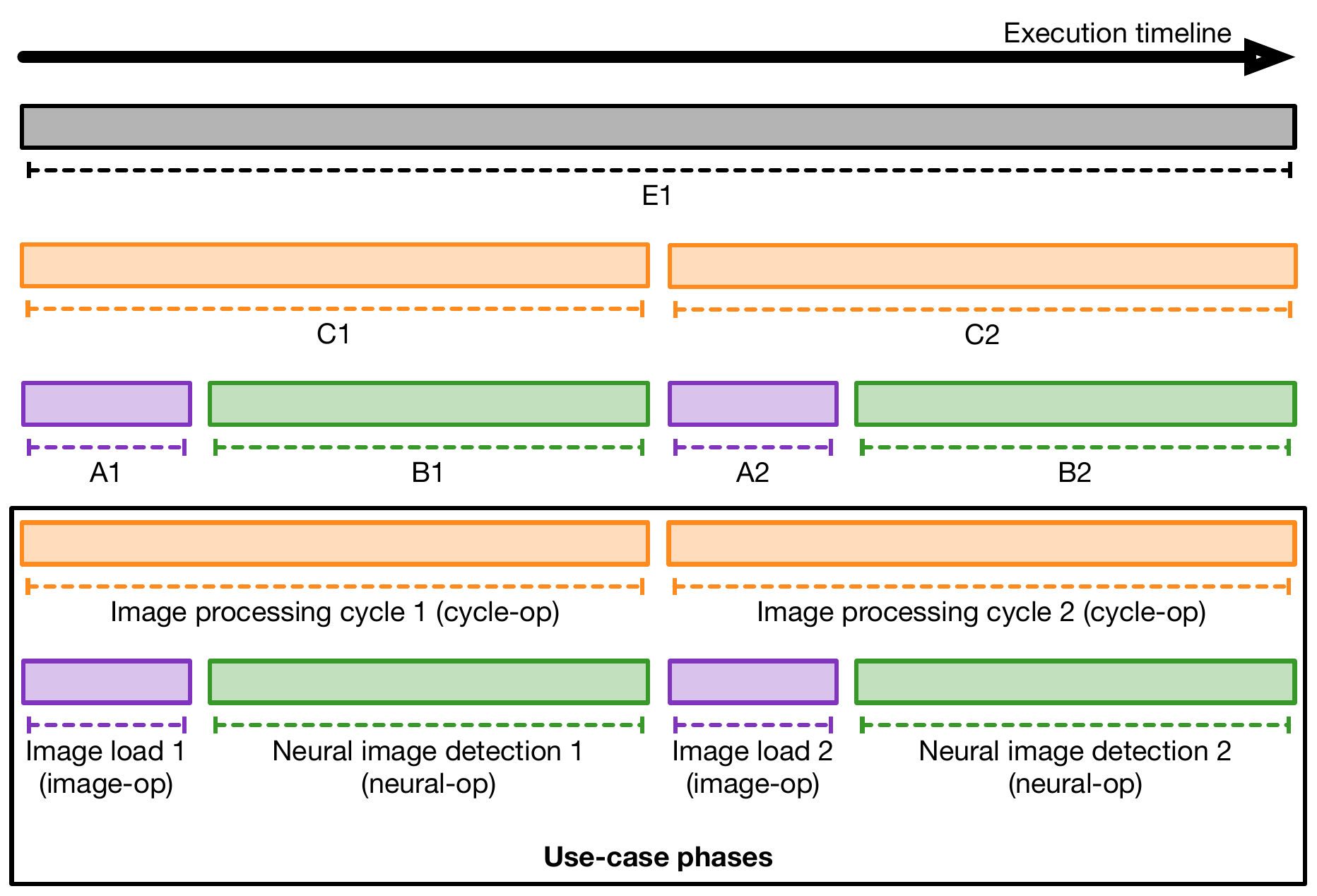}
	\caption{Different data compartmentalisation granularities within an execution timeline, visualising repeated phase types during consecutive rounds of tasks/sub-tasks, plus the phases considered for processing consecutive image data.}
	\label{fig:execution_phases}
\end{figure}

One of the preprocessing steps governing the dataset formation for ML model training is the compartmentalisation of data according to execution phases. Execution phases are segments of monitoring data, reflecting the behaviour of the system under scrutiny per individual task or subtask during an execution. In other words, the time-series data from the start till the stop signal is associated with the respective task. In this fashion, the data can be \emph{cut} in individual phases. \Cref{fig:execution_phases} depicts the concept and provides the phases considered in our demonstration use-case.

\subsection{Emulating data corruptions}
\CPSLint is capable of both sanitising and compartmentalising time-series data. To be able to systematically evaluate and demonstrate these functionalities, a series of corrupt data examples are devised. This is done through Python scripts manipulating reference data, i.e., correct data, and emulating known corruption patterns. The considered reference data is collected from an embedded platform with repetitive sequential activity, as described by Odyurt et al.~\cite{Odyurt:2021:PPFT}.

The corruptions are applied to the reference data in non-overlapping blocks of rows, in this case 10-row blocks. The total corruption rate is 0.5\%, which is a good amount considering how extensiveness the reference data is. Note that while not affecting \CPSLint functionality, the block size and corruption rate are configurable parameters. Blocks are selected at random unless the emulated corruption is targeted. Corruptions can be applied individually, or in combination. The following types are supported:

\paragraph*{Data type mismatch}
Numeric fields are injected with invalid characters, e.g., letters or special symbols, rendering them non-parsable. In some cases, the Universal Asynchronous Receiver/Transmitter (UART) identifier field is also corrupted, simulating noisy channel logs.

\paragraph*{Data type mismatch with targeted UART}
Similar to the previous corruption, but biased towards affecting rows associated with a specified UART identifier. This enables testing of selective corruption of device-specific data. UART data fields are especially important for steering the targeted solution and as such, it is important to have cases with noisy UART messages.

\paragraph*{Out-of-bounds values}
Numeric fields are replaced with extreme constants, e.g., \enquote{99\,999.999}, that are far outside expected measurement ranges. Depending on configuration, this may affect all numeric columns or a random subset within each block.

\paragraph*{Out-of-order rows with reliable timestamps}
Rows in a block are randomly permuted, but their original timestamps are preserved. This emulates reordering of messages while maintaining temporal information.

\paragraph*{Out-of-order rows with unreliable timestamps}
Rows are shuffled and new, monotonically increasing timestamps are generated for the block. This emulates cases where both order and timing data are compromised during collection.

\paragraph*{Missing fields}
A randomly chosen numeric column is blanked out across all rows in a block, representing systematic loss of a sensor channel.

\paragraph*{Missing rows}
Entire blocks of rows are deleted, optionally biased toward containing a specified UART identifier. This models partial loss of data segments, such as dropped packets or truncated logs.

\paragraph*{Misplaced end-of-line markers}
Rows within a block are concatenated with their successors, simulating the effect of missing or corrupted line delimiters. This results in malformed records and broken row alignment, resembling common parsing errors in raw log files.

\section{Using \CPSLint}
\label{sec:application}

\subsection{Setup}
To set up \CPSLint in a workspace, the DSL needs to be configured with a set of parameters. These include location of the input CSV files, the desired output folder, a command to invoke Python, and which pipeline to use for execution of the specification. The parameters are to be defined in a YAML file in the folder where the \CPSLint definitions are located. An example of such a configuration file is given in \Cref{lst:config}.
\begin{figure}[h!]
    \centering
\begin{lstlisting}[ 
        caption={The config.yaml}, 
        label=lst:config, 
        language=yaml
    ]
inputFolder : ~/data/input
outputFolder : ~/data/output
pythonCommand : python
runWith: interpreter # Current options: interpreter, compiler
\end{lstlisting}
\end{figure}

As can be seen in \Cref{lst:config}, \CPSLint allows for the use of multiple back-ends: a compiler pipeline, meant for generating well performing idiomatic Python code, and an interpreter pipeline, a back-end meant for debugging and diagnostic purposes at the cost of performance. A more detailed overview of these pipelines is given in \Cref{sec:implementation}.  

Regardless of which back-end pipeline is chosen, the user can use the language in one of two ways: through the Rascal REPL, or through the Language Server Protocol (LSP) interface provided by Rascal. Accessing \CPSLint through the Rascal REPL allows the use of the language without the overhead of an IDE. When considering this option, the \texttt{Runner} module can be imported to run the \CPSLint programs through calling the \texttt{run} function from the REPL, by giving the location of the program to be executed as an argument. An example of this usage pattern is given in \Cref{lst:REPL}. 
\begin{figure}[h!]
    \centering
\begin{lstlisting}[
        caption={An example showing how to run \CPSLint specifications through the Rascal REPL.},
        language={},
        label=lst:REPL,
        breaklines=true,
        breakatwhitespace=false,
        columns=fullflexible,
        keepspaces=true
    ]
rascal>import Runner;
ok
rascal>run(|project://cps/Input/scripts/input.cps|);
\end{lstlisting}
\end{figure}

\begin{figure*}[htbp]
	\centering
	\includegraphics[width=0.8\linewidth]{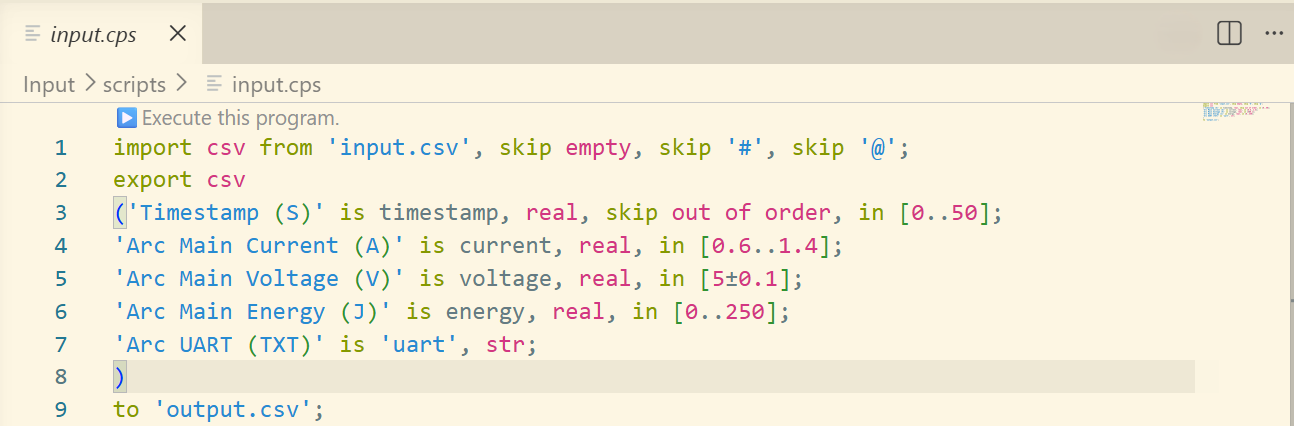}
	\caption{\CPSLint language features running in Visual Studio Code.}
	\label{fig:IDE}
\end{figure*}

Using the LSP implementation gives the user Visual Studio Code support, bringing \CPSLint language features to the editor. These features can be enabled by importing the \texttt{CPSLintLanguageServer} module through the Rascal REPL and running the \texttt{main} function, which in turn registers the \CPSLint language and makes the editor ready for use. In \Cref{fig:IDE}, we illustrate what the IDE environment looks like. As can be seen, we provide syntax highlighting for the language and a way to run the current program.

\subsection{Main features}
The current implementation of the \CPSLint language provides the following core feature set:

\paragraph*{Data inspection and script generation} 
This feature is meant as a starting point when working with a new dataset using \CPSLint. Here the user can write a one-line command such as \lstinline|inspect csv from 'input.csv';|, which instructs \CPSLint to produce a specification based on the given CSV file. Users can then use this to continue defining how the CSV in question should be processed.

\paragraph*{Data corruption remediation} 
A large chunk of \CPSLint's functionality revolves around the remediation of corrupt data. This includes features such as enforcing datatypes on columns to ensure that only values of an expected type are present. It can also set valid ranges for numeric columns, such that faulty readings are removed from the dataset. Additionally, it may filter unwanted substrings out of the data, ensuring that any unexpected characters that appear due to, for example, lossy channels are removed.

\paragraph*{Data imputation}
\CPSLint also allows for the imputation of missing data. The interpolation methods the language offers range from simple ones, such as using the mean or median of a column, or forward or backfilling the column with known values, to more involved ones, such as polynomial interpolation.

\paragraph*{Data compartmentalisation}
\CPSLint supports data compartmentalisation through the \lstinline|export| command. Using this command, the user can specify based on which column they would like to split a file into smaller ones. For example, \newline
\lstinline|export csv  ... to 'output#.csv' cut when 'UART' is 'image loader'~;| \newline
instructs \CPSLint to create multiple output CSV files, each covering a unique instance where image loading happens.

\subsection{Example programs}
\label{subsec:exprogs}
Here, we will go over a few representative input programs for \CPSLint and explain them in detail.

\paragraph{Example program 1: \texttt{out\_of\_bounds.cps}}
This program is representative of a case where it can be assumed that the input CSV has correct column types, but contains rows in which column values lie outside the defined valid ranges.
\begin{figure}[h!]
    \centering
\begin{lstlisting}[
        caption={\CPSLint code for Example Program 1}, 
        label=lst:ex1
    ]
import csv from 'out_of_bounds.csv',     skip empty;
export csv
(
  'Timestamp (S)' is timestamp, real,      in (0..120];
  'Arc Main Current (A)' is current, real,  in [0.4..1.0);
  'Arc Main Voltage (V)' is voltage, real,  in [5±0.6];
  'Arc Main Energy (J)' is energy, real,    in (0..400);
  'Arc UART (TXT)' is 'uart';
)
to 'out_of_bounds_fixed.csv';
\end{lstlisting}
    \vspace{-1em}
\end{figure}

In \Cref{lst:ex1}, we show a program where first a CSV is imported through the \lstinline|import| command, then empty lines are removed through the \lstinline|skip empty| row filter; row filters are applied at the row level across the file. The second command in the file is the \lstinline|export| command, here the columns found in the import file are mapped to rows in the output file through the \lstinline|is| keyword. Then for all numeric columns, the DSL is instructed to view them as \lstinline|real| values. Finally, the \lstinline|in| keyword is used to indicate a valid numeric range for each column before exporting the CSV to the output file. Cells where the data is outside the defined valid range of a column are emptied without affecting the rest of the row.

When this program is run using the \CPSLint compiler, the code in \Cref{lst:ex1py} is generated as an intermediate artefact; when executed, this Python code produces the output CSV.

\paragraph{Example program 2: \texttt{datatype\_mismatch.cps}}
The second program is representative to one that would be written when there are unexpected types encountered in columns where a certain type is expected. In this case there are string values in numeric columns that need to be removed. 
\begin{figure}[h!]
    \centering
\begin{lstlisting}[
        caption={\CPSLint code for Example Program 2}, 
        label=lst:ex2
    ]
import csv from 'datatype_mismatch.csv', skip empty, skip ~'#'~, skip ~'@'~, skip ~'$'~, skip ~'abc'~;
export csv
(
  'Timestamp (S)' is timestamp, real;
  'Arc Main Current (A)' is current, real, impute linear interpolation;
  'Arc Main Voltage (V)' is voltage, real, impute linear interpolation;
  'Arc Main Energy (J)' is energy, real, impute linear interpolation;
  'Arc UART (TXT)' is 'uart';
)
to 'datatype_mismatch_fixed.csv';
\end{lstlisting}
    \vspace{-1em}
\end{figure}

The \lstinline|import| command in \Cref{lst:ex2} starts off similar to the one in the first example program, however here there are also file wide substring filters defined by the \lstinline|skip| column filters in the command. The tildes in the \lstinline|skip| filters signify if a string should be skipped, if it starts with a substring (tilde after the substring), if it ends with a substring (tilde before the substring), or if it contains the substring (substring between tildes). In other words, the tilde notation differentiates between substring filtering (if tildes are present) and exact string filtering (if tildes are absent). The \lstinline|export| command starts off by enforcing the appropriate datatypes on the columns. Furthermore, the \lstinline|export| command fills any cells that might have been emptied by previous operations, or were already empty in the input CSV, via the \lstinline|impute| keyword, using \lstinline|linear interpolation| as an imputation strategy.

\begin{figure*}[htbp]
    \centering
\begin{lstlisting}[
        caption={Python code generated by running Example Program 1 using the \CPSLint compiler}, 
        language=Python, 
        label=lst:ex1py
    ]
# Generated by CPSlint
# Input: out_of_bounds.cps

# Import libraries
...

# Auxiliary functions
...

def call_cpslint():
    # Read input CSV
    input_path = os.path.join("~/data/input", "out_of_bounds.csv")
    df_in = pd.read_csv(input_path)
    # Skip empty rows
    df_in = df_in.replace(r'^\s*$', pd.NA, regex=True).dropna(how="all")
    # Create output DataFrame
    df_out = pd.DataFrame()
    # Copy column "Timestamp (S)" as "timestamp" in the output DataFrame 
    df_out["timestamp"] = df_in["Timestamp (S)"]
    # Copy column "Arc Main Current (A)" as "current" in the output DataFrame 
    df_out["current"] = df_in["Arc Main Current (A)"]
    # Copy column "Arc Main Voltage (V)" as "voltage" in the output DataFrame 
    df_out["voltage"] = df_in["Arc Main Voltage (V)"]
    # Copy column "Arc Main Energy (J)" as "energy" in the output DataFrame 
    df_out["energy"] = df_in["Arc Main Energy (J)"]
    # Copy column "Arc UART (TXT)" as "uart" in the output DataFrame 
    df_out["uart"] = df_in["Arc UART (TXT)"]
    # Cast values of timestamp to float if possible, set to NaN otherwise 
    df_out["timestamp"] = pd.to_numeric(df_out["timestamp"], errors="coerce").astype(float)
    # Emptying cells in timestamp that are not in the range between 0. and 120., with inclusive set to right
    df_out.loc[~df_out["timestamp"].between(0., 120., inclusive="right"), "timestamp"] = np.nan
    # Cast values of current to float if possible, set to NaN otherwise 
    df_out["current"] = pd.to_numeric(df_out["current"], errors="coerce").astype(float)
    # Emptying cells in current that are not in the range between 0.4 and 1.0, with inclusive set to left
    df_out.loc[~df_out["current"].between(0.4, 1.0, inclusive="left"), "current"] = np.nan
    # Cast values of voltage to float if possible, set to NaN otherwise 
    df_out["voltage"] = pd.to_numeric(df_out["voltage"], errors="coerce").astype(float)
    # Emptying cells in voltage that are not in the range between 4.4 and 5.6, with inclusive set to both
    df_out.loc[~df_out["voltage"].between(4.4, 5.6, inclusive="both"), "voltage"] = np.nan
    # Cast values of energy to float if possible, set to NaN otherwise 
    df_out["energy"] = pd.to_numeric(df_out["energy"], errors="coerce").astype(float)
    # Emptying cells in energy that are not in the range between 0. and 400., with inclusive set to neither
    df_out.loc[~df_out["energy"].between(0., 400., inclusive="neither"), "energy"] = np.nan
    # Write Output CSV
    output_path = os.path.join("~/data/output", "out_of_bounds_fixed.csv")
    df_out.to_csv(output_path, index=False)
call_cpslint()
\end{lstlisting}
\end{figure*}

The Python code generated by the \CPSLint compiler can be found in \Cref{lst:ex2py}. It should be noted that most interpolation methods supported by Pandas rely on SciPy being installed.

\subsection{Planned features}
Future iterations of \CPSLint are planned to bring features that would better serve the needs of the domain for which it is built. The declarative nature of the syntax of \CPSLint makes the language relatively easy to extend at the syntax level, with new features requiring only minor edits of the grammar. The main additions would be to introduce the concept of domain-specific units (Volt, Watt, mAh) as native data types. This would enable us to perform safe and simple type conversions on columns. Another extension improving applicability, is to support more data source formats, e.g., HDF5 files or time-series databases. Future iterations of \CPSLint could support compartmentalisation based on numerical columns, where, for instance, time-series data is cut according to peaks, valleys, and plateaus in these values. This is useful in cases where the data cannot be cut based on clear logged indicators due to their absence, unreliability, or corrupt state. 

Another extension that would make the language more broadly applicable is the support for custom Python libraries within the compiler pipeline. These would contain functions taking either a \texttt{DataFrame} (for operations at the table level) or a \texttt{Series} (for operations at the column level) as their input. In this way, users can add their own case-specific algorithms to preprocess data using \CPSLint. An example of how such an extension could look like is depicted in \Cref{lst:pylibs}.

\begin{figure}[h!]
    \centering
\begin{lstlisting}[
        caption={Syntax for using external libraries in \CPSLint.}, 
        label={lst:pylibs}
    ]
using <library.py>;
import csv from 'input.csv', skip empty, skip '#', skip '@', perform <tableOperation>;
export csv
(
  ...
  'Arc Main Current (A)' is current, real, in [0.6..1.4], perform <rowOperation>;
  ...
)
to 'output.csv';
\end{lstlisting}
\end{figure}

\begin{figure*}[htbp]
    \centering
\begin{lstlisting}[
        caption={Python code generated by running Example Program 2 using the \CPSLint compiler}, 
        language=Python, 
        label=lst:ex2py
    ]
# Generated by CPSlint
# Input: datatype_mismatch.cps

# Import libraries
...

# Auxiliary functions
...

def call_cpslint():
    # Read input CSV
    input_path = os.path.join("~/data/input", "datatype_mismatch.csv")
    df_in = pd.read_csv(input_path)
    # Blank out values that contain #
    df_in = df_in.replace(r'^.*' + re.escape("#") + r'.*$', '', regex=True)
    # Blank out values that contain @
    df_in = df_in.replace(r'^.*' + re.escape("@") + r'.*$', '', regex=True)
    # Blank out values that contain $
    df_in = df_in.replace(r'^.*' + re.escape("$") + r'.*$', '', regex=True)
    # Blank out values that contain abc
    df_in = df_in.replace(r'^.*' + re.escape("abc") + r'.*$', '', regex=True)
    # Skip empty rows
    df_in = df_in.replace(r'^\s*$', pd.NA, regex=True).dropna(how="all")
    # Create output DataFrame
    df_out = pd.DataFrame()
    # Copy column "Timestamp (S)" as "timestamp" in the output DataFrame 
    df_out["timestamp"] = df_in["Timestamp (S)"]
    # Copy column "Arc Main Current (A)" as "current" in the output DataFrame 
    df_out["current"] = df_in["Arc Main Current (A)"]
    # Copy column "Arc Main Voltage (V)" as "voltage" in the output DataFrame 
    df_out["voltage"] = df_in["Arc Main Voltage (V)"]
    # Copy column "Arc Main Energy (J)" as "energy" in the output DataFrame 
    df_out["energy"] = df_in["Arc Main Energy (J)"]
    # Copy column "Arc UART (TXT)" as "uart" in the output DataFrame 
    df_out["uart"] = df_in["Arc UART (TXT)"]
    # Cast values of timestamp to float if possible, set to NaN otherwise 
    df_out["timestamp"] = pd.to_numeric(df_out["timestamp"], errors="coerce").astype(float)
    # Cast values of current to float if possible, set to NaN otherwise 
    df_out["current"] = pd.to_numeric(df_out["current"], errors="coerce").astype(float)
    # Impute with stategy: empty cells in current are replaced using interpolation with the linear method
    df_out["current"] = df_out["current"].interpolate(method="linear")
    # Cast values of voltage to float if possible, set to NaN otherwise 
    df_out["voltage"] = pd.to_numeric(df_out["voltage"], errors="coerce").astype(float)
    # Impute with stategy: empty cells in voltage are replaced using interpolation with the linear method
    df_out["voltage"] = df_out["voltage"].interpolate(method="linear")
    # Cast values of energy to float if possible, set to NaN otherwise 
    df_out["energy"] = pd.to_numeric(df_out["energy"], errors="coerce").astype(float)
    # Impute with stategy: empty cells in energy are replaced using interpolation with the linear method
    df_out["energy"] = df_out["energy"].interpolate(method="linear")
    # Write Output CSV
    output_path = os.path.join("~/data/output", "datatype_mismatch_fixed.csv")
    df_out.to_csv(output_path, index=False)
call_cpslint()
\end{lstlisting}
\end{figure*}

\section{Design and implementation}
\label{sec:implementation}
\CPSLint is implemented using the Rascal meta programming language~\cite{Klint:2009:RDSL, Klint:2011:EMPR}, a \enquote{one-stop-shop} language workbench for rapid implementation of DSLs. The language offers its users a shared front-end and two back-ends: the \emph{compiler}, which generates human-readable, idiomatic Python code that performs the tasks described by the \CPSLint input program; and the direct interpreter in Rascal, providing granular insight into the data sanitisation process in the form of logs and intermediate CSV files. In this section, we go over the internals of \CPSLint.

\subsection{Language front-end}
The syntax of \CPSLint has been kept relatively simple, with the language only knowing three actions: \lstinline|inspect| for initial analysis of the data, \lstinline|import| for reading, and \lstinline|export| for writing. The syntax of the language was created during design sessions with domain experts, with the goal of making a language that is familiar, sufficiently expressive, but not overly bloated.
The entire abstract syntax definition, illustrating the simplicity of the DSL, is contained in \Cref{lst:grammar}.
\begin{figure*}[htbp]
    \centering
\begin{lstlisting}[
        caption={The abstract syntax definition of \CPSLint.}, 
        language=Rascal, 
        label=lst:grammar
    ]
module syntaxes::Abstract

alias Script = list[Action];
alias Mapping = tuple[str from, str to, list[Filter] filters];
alias Mappings = lrel[str from, str to, list[Filter] filters];
alias Border = tuple[str name, Stringish val];

data Action
    = ImportAction(DataFormat format, list[Filter] filters, str filename)
    | ExportAction(DataFormat format, list[Mapping] mappings, str filename, Border border)
    | InspectAction(DataFormat format, str filename)
    | Inaction(str message)
    ;

data DataFormat = CSV();

data Stringish
    = JustString(str text)
    | RegExp(str regex)
    | StartsWith(str begin)
    | EndsWith(str end)
    | Contains(str middle)
    ;

data Filter
    = SkipEmptyRows()
    | SkipEmptyColumns() 
    | SkipOutOfOrder()
    | SkipIgnored(Stringish what)
    | InRange(real low, real high, bool includeLow, bool includeHigh)
    | EnforceType(SupportedType t)
    | EnforceOrder()
    | Impute(Strategy s)
    | IsCritical()
    | Unique()
    ;

data SupportedType = Natural() | Integer() | Real() | String() | Boolean() | UART();

data Strategy
    = Last()
    | Next()
    | Mean()
    | Median()
    | Constant(str c)
    | Interpolation(str method)
    ;
\end{lstlisting}
\end{figure*}

The parsing infrastructure is rather straightforward: we first parse the \CPSLint programs as a concrete syntax tree, then implode it into a more workable abstract syntax tree. The abstract syntax tree is then passed to the selected back-end for the current run.

\subsection{Ad hoc Python sanitisation}
\begin{figure*}[htbp]
    \centering
    \begin{subfigure}{\linewidth}
    	\centering
	    \includegraphics[width=0.75\linewidth]{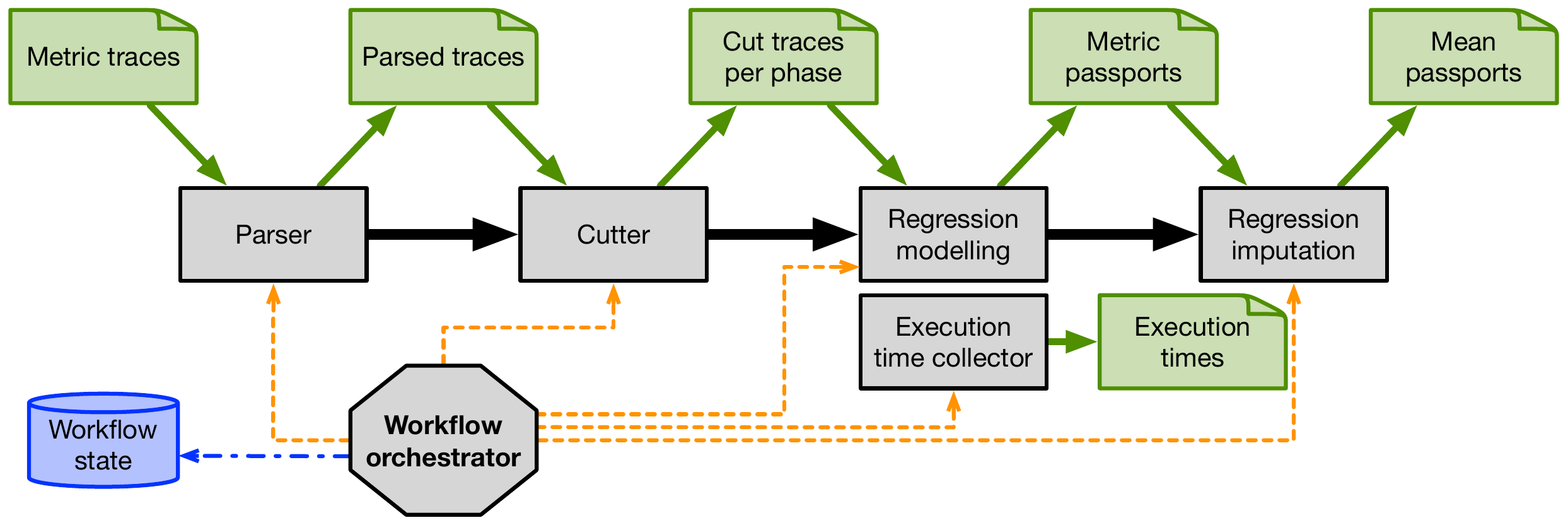}
	    \caption{Solution workflow processing normal system data, intended for traditional ML dataset formation.}
	    \label{fig:solution_flow_traditional_ml_normal}
    \end{subfigure}
    \vspace{0.5em}
    \begin{subfigure}{\linewidth}
    	\centering
    	\includegraphics[width=\linewidth]{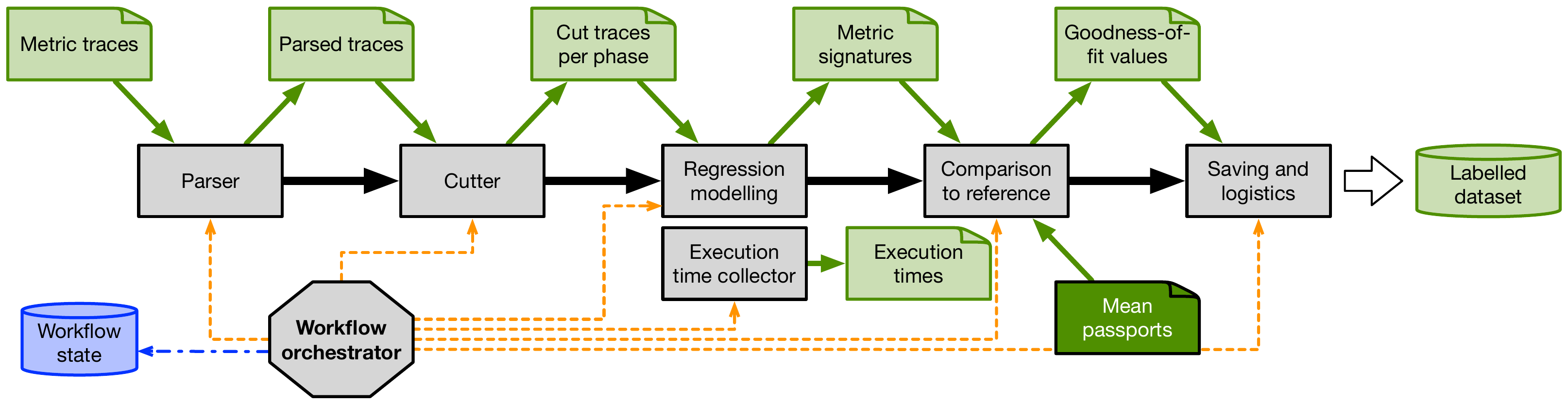}
		\caption{Solution workflow processing anomalous system data, intended for traditional ML dataset formation.}
		\label{fig:solution_flow_traditional_ml_anomalous}
    \end{subfigure}
    \vspace{0.5em}
    \begin{subfigure}{\linewidth}
    	\centering
    	\includegraphics[width=0.8\linewidth]{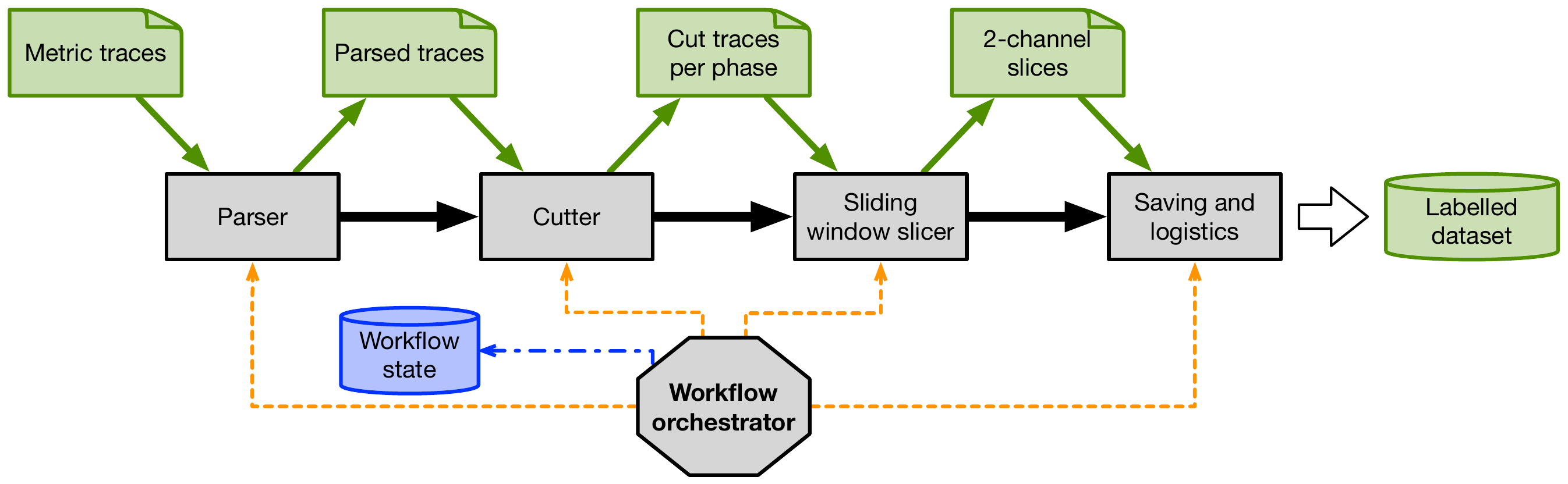}
		\caption{Solution workflow processing system data, intended for CNN dataset formation.}
		\label{fig:solution_flow_cnn}
    \end{subfigure}
	\caption{Different workflows involving parsing and cutting of industrial CPS monitoring data: Traditional ML model training dataset formation flow handling a) \enquote{Normal} data and b) \enquote{Anomalous} data; c) CNN model training dataset formation flow. Note the presence of parsing and cutting in all such workflows as fundamental and repeating data processing stages.}
	\label{fig:solution_flows}
\end{figure*}

As both our inspiration and the actual use-case in which \CPSLint is a suitable replacement for some of the stages, a Python project~\cite{Odyurt:2021:CAMA} containing multiple configurable workflows, composed of sequential pipeline stages, can be presented. Details of the monitoring data format and fields relevant to this use-case have been elaborated in \Cref{sec:data}. The project provides alternative anomaly detection and identification solutions for industrial CPS, based on traditional ML algorithms and CNN deep neural networks. \Cref{fig:solution_flows} covers three of these workflows, which prepare and process data for ML dataset formation. The follow-up workflows using these datasets and performing actual model training or inference, are not included, as there is no need for \CPSLint's functionality at those stages.

In short, \Cref{fig:solution_flow_traditional_ml_normal,fig:solution_flow_traditional_ml_anomalous} depict the data processing steps for processing data representing normal and anomalous system activity for traditional ML training. On the other hand, \Cref{fig:solution_flow_cnn} does the same, but for a CNN model design. As it can be seen, all three workflows involve \emph{parsing} and \emph{cutting} (compartmentalisation) of the data. These tasks match the functionality provided by \CPSLint, i.e., machine trace ingestion, sanitisation, and compartmentalisation. As these steps are commonly present not only in these workflows, but in general most workflows processing industrial CPS data, the need for reusable tools such as \CPSLint is apparent.

\subsection{The compiler back-end}
The compiler back-end takes the parsed abstract syntax tree and generates the corresponding Python code from it. This code is then run, producing the processed CSV file.

The compiler generates Python code that performs the operations described in the \CPSLint input. The generated code leverages existing Python data science libraries such as Pandas, NumPy, and SciPy, enabling the compiler to produce high-performance, relatively compact code. Each code fragment generated by the compiler is preceded by a short description of its purpose. This can be seen in the example programs and their generated code shown in \Cref{subsec:exprogs}. Users can thus relate the generated code to the \CPSLint specification and make changes to either the \CPSLint program or the generated Python code if the compiled code does not match their expectations.

\begin{figure*}[htbp]
	\centering
    \includegraphics[width=0.90\linewidth]{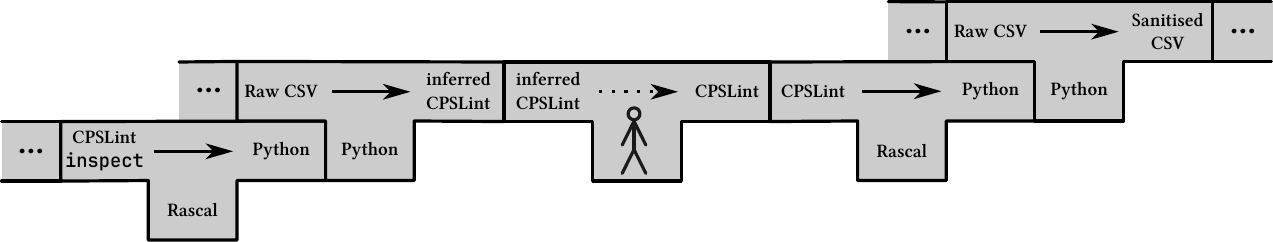}
	\caption{A tombstone diagram of the \CPSLint compiler pipeline. Activities flow rightwards, with inspection and inferring of the data structure happening on the left side, followed by Python code generation to deliver the executable code operating on the actual machine trace. Ellipses indicate where the normal data processing flow connects. Note the role of a domain expert refining the inferred \CPSLint specification.}
	\label{fig:cpslint_pipeline}
\end{figure*}

The pipeline for this back-end is illustrated in \Cref{fig:cpslint_pipeline}. The figure starts with an \lstinline|inspect| script, which generates a script describing the input dataset in its unprocessed form. The compiler generates a Python script that in turn generates a baseline \CPSLint specification for the dataset. This is done with the help of the Pandas library, which is used to infer column names and types. The domain expert then refines the baseline script by adding how \CPSLint should sanitise the dataset. This refined input is then compiled into a Python script, which is in turn run to output a sanitised version of the dataset.

\subsection{The interpreter back-end}
The interpreter back-end directly interprets the \CPSLint code, performing the operations defined in the input using Rascal itself and thus forgoing compilation to Python entirely. This method of running \CPSLint relies on our runtime library written in Rascal and thus cannot make use of the more mature, well-performing data science libraries available for Python, and therefore suffers a significant performance penalty. What this back-end does offer, however, is greater insight into the individual steps \CPSLint takes while executing the input script. It does so by outputting an intermediate CSV after each operation, and by producing a timestamped log detailing the steps taken to perform the data sanitisation at runtime. An example of such a log file is given in \Cref{lst:log}.
\begin{figure*}[htbp]
    \centering
\begin{lstlisting}[
        caption={A truncated example of the log produced by the interpreter back-end}, 
        language={}, 
        literate={}, 
        label=lst:log
    ]
2026-04-08 15:25:34.597 | [IO] Created intermediate CSV for Import action ~/data/output/intermediate/import_filters_applied.csv
2026-04-08 15:25:34.627 | [ImportFilter] Skipping empty rows
2026-04-08 15:25:37.019 | [RuntimeLib] Saved new intermediate csv after applying skipEmptyRows at ~/data/output/intermediate/import_filters_applied_skipEmptyRows.csv
2026-04-08 15:25:37.053 | [RuntimeLib] Updated ~/data/output/intermediate/import_filters_applied.csv after applying skipEmptyRows
2026-04-08 15:25:37.054 | [ImportFilter] Skipped empty rows
2026-04-08 15:25:37.055 | [IO] CSV used for Export: ~/data/output/intermediate/import_filters_applied.csv
...
2026-04-08 15:31:28.762 | [ColumnFilters] Enforced type real on column "energy"
2026-04-08 15:31:28.762 | [ColumnFilters] Emptying cells in energy that are not in the range between 0. and 400., with inclusive set to neither
2026-04-08 15:32:35.563 | [RuntimeLib] Wrote file where column "energy" only has values in range between 0. and 400., with inclusive set to neither to ~/data/output/intermediate/export_filters_applied_energy_range_applied.csv
2026-04-08 15:32:35.632 | [RuntimeLib] Updated ~/data/output/intermediate/export_filters_applied.csv after enforcing range (0. and 400., inclusive: neither) on column "energy"
2026-04-08 15:32:35.633 | [ColumnFilters] emptied cells in energy that are not in the range between 0. and 400., with inclusive set to neither
2026-04-08 15:32:35.633 | [IO] Applied column filters
2026-04-08 15:32:35.640 | [IO] Saved final result to ~/data/output/out_of_bounds_fixed.csv
\end{lstlisting}
\end{figure*}

\begin{figure}[htbp]
	\centering
    \includegraphics[width=\linewidth]{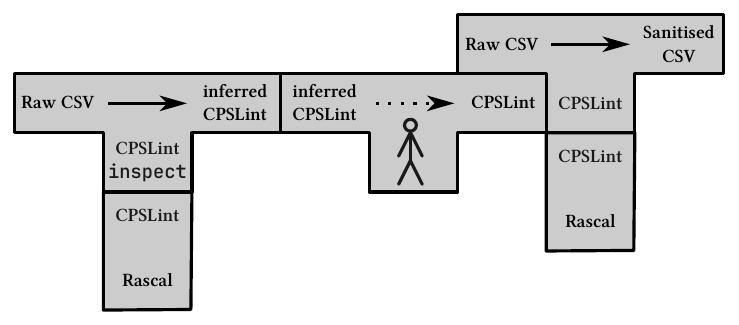}
	\caption{A tombstone diagram of the \CPSLint interpreter: the presence of the interpreter (vertical block) of \CPSLint written in Rascal allows us to see \CPSLint specifications as transformations from raw CSV to its sanitised form. }
	\label{fig:cpslint_pipeline_interpreter}
\end{figure}

The tombstone diagram for using \CPSLint with the interpreter back-end is shown in \Cref{fig:cpslint_pipeline_interpreter}. Like with the compiler back-end the process starts with providing \CPSLint with an \lstinline|inspect| script, which again generates a baseline \CPSLint script. This back-end uses \lstinline[language = Rascal]|lang::csv| from the Rascal standard library for type inference of the column names and types, thereby matching the functionality from the \lstinline|inspect| implementation from the compiler. The domain expert again refines the baseline script and runs this through the interpreter, which in this case generates a sanitised CSV as its main output. The interpreter writes the log alongside the main output, while creating intermediate files after each step the interpreter performs during the execution.

\subsection{Integration into pipelines}
\begin{figure}[htbp]
    \centering
    \begin{subfigure}{\linewidth}
    	\centering
	    \includegraphics[width=\linewidth]{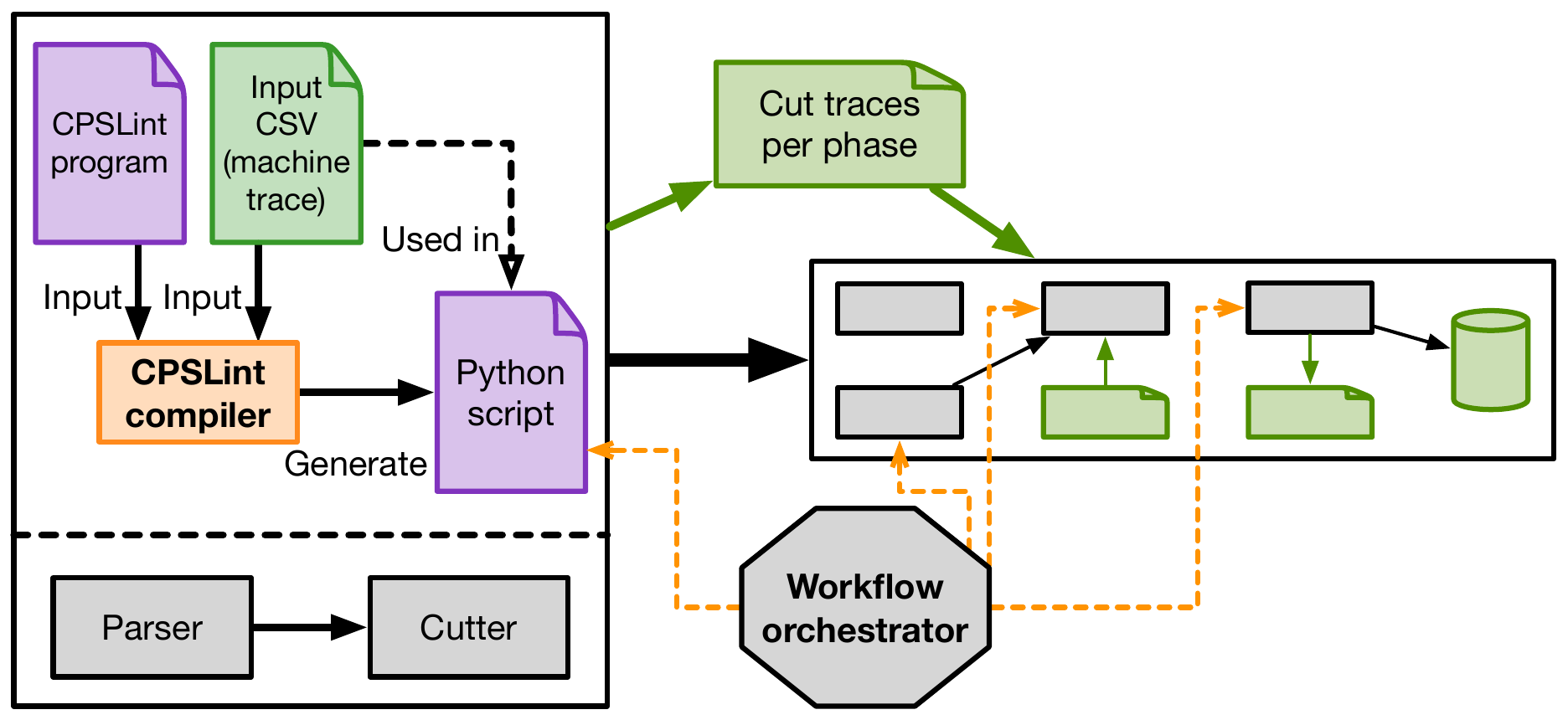}
	    \caption{\CPSLint flow in compiler mode.}
	    \label{fig:cpslint_compiler_flow}
    \end{subfigure}
    \vspace{0.5em}
    \begin{subfigure}{\linewidth}
    	\centering
    	\includegraphics[width=\linewidth]{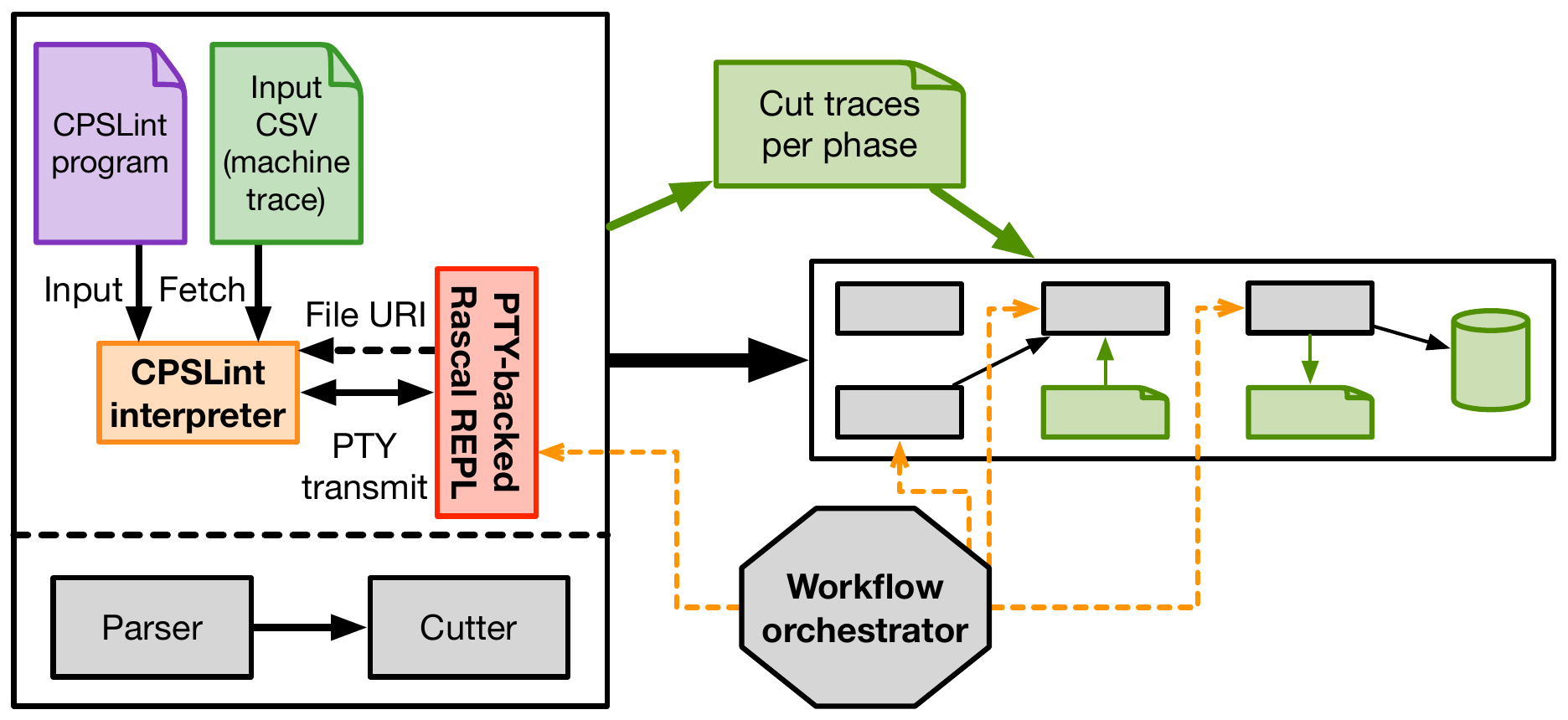}
		\caption{\CPSLint flow in interpreter mode.}
		\label{fig:cpslint_interpreter_flow}
    \end{subfigure}
	\caption{Integration options considering the two modes of operation offered by \CPSLint, i.e., a) compiler and b) interpreter modes.}
	\label{fig:integrated_flows}
\end{figure}

Considering the flows from \Cref{fig:solution_flows}, parsing and cutting stages in any of these, or any other time-series processing flow for that matter, can be replaced by the functionality provided by \CPSLint. As described at the start of this section, \CPSLint support two modes of operation: as a compiler and as an interpreter. Diagrams in \Cref{fig:integrated_flows} present how the replacement and integration of \CPSLint can be conceived.

As described in \Cref{sec:application}, the interpreter mode is primarily intended for debugging and diagnostic purposes. The computational cost incurred and the I/O-heavy design will render this mode inadequate for production. Regardless, a systematic integration that can be driven with an orchestrator is shown in \Cref{fig:cpslint_interpreter_flow}, mainly through a pseudoterminal (PTY) and a Read-Eval-Print Loop (REPL) interface.

The compiler mode on the other hand, could be seamlessly integrated with any workflow, as the orchestrator only has to drive Python scripts generated by \CPSLint (\Cref{fig:cpslint_compiler_flow}).

\section{Related work}
\label{sec:related_work}
\CPSLint was implemented in Rascal, which is just one of the language workbenches available today~\cite{Erdweg:2015:ECLW}. What is specific to Rascal is that it combines the role of meta-programming language and language workbench, with explicit support for source-code analysis and transformation~\cite{Klint:2009:RDSL}. Its design brings together grammar definition, algebraic data types, pattern matching, tree traversal, and source-location-aware manipulation in a single executable environment~\cite{Klint:2011:EMPR}. Among contemporary language workbenches, this places Rascal on the textual side, but with a particularly strong emphasis on programmable analyses and transformations~\cite{Erdweg:2015:ECLW}.

Among the tools for data wrangling, sanitisation and compartmentalisation, we recognise the following notable examples, each providing various sets of features.

\emph{GNU datamash}~\cite{GNU:2025:Datamash} is a command-line tool that performs basic numeric, textual, and statistical operations on textual data. While it is not a fully-fledged DSL, it is suitable for handling basic manipulation and analysis on the input. Compared to \CPSLint, it provides general-purpose command-line operations rather than a domain-specific notation for sanitising and compartmentalising industrial time-series data.

\emph{Lisp Query Notation}~\cite{Hoff:2024:LQND} is an embedded DSL in Common Lisp that can operate on CSV data as well as structured formats such as JSON. It allows users to query and manipulate textual data both as a DSL and as a command-line tool. Being embedded in Common Lisp enables users to extend its functionality and makes it possible to fall back on Lisp when the DSL is limiting for certain operations. It is more host-language-extensible and in general broader than \CPSLint with its small declarative vocabulary tailored to sanitisation, imputation, and compartmentalisation.

\emph{DescribeML}~\cite{Giner-Miguelez:2023:DSLD} is a DSL designed to precisely describe machine learning datasets in terms of their structure, provenance, and social concerns, with the aim of providing a method for dataset documentation based on their characteristics, thereby helping data scientists select the appropriate dataset for a given project. DescribeML is implemented using Langium and has been published as a Visual Studio Code extension. Unlike \CPSLint, it targets dataset description and documentation rather than executable preprocessing of raw traces.

\emph{RADAR}~\cite{Heine:2020:DADQ} is a DSL built for data quality monitoring. RADAR's capabilities include simple checks like making sure there are no null values, data integrity checks and statistical checks like conformance of the data to a certain  distribution. This DSL is made up of 3 types of definitions; sources (referring to the input data), checks (which act like functions that accept parameters), and actions (which apply the checks on data from a source and define what should be reported for the resulting values). In contrast to \CPSLint, RADAR emphasises monitoring and reporting of data quality rather than transforming corrupted data into sanitised output datasets.  

\emph{Lavoisier}~\cite{DeLaVega:2020:LDIL} is a DSL that helps with creation of datasets that conform to the format accepted by data mining algorithms. It is meant to reduce script size compared to accomplishing the same result using other ways to prepare data such as SQL and Pandas. Its focus is broader dataset preparation for data mining without the CPS focus. \emph{Papin}~\cite{Sal:2024:DSLA} is a spin-off DSL of Lavoisier that specialises further and makes it compatible with fishbone diagrams, a notation used by industrial engineers to model cause and effect. For the Papin pipeline the authors introduce a new fishbone diagram kind called data-oriented fishbone diagrams which combines domain data (defined using Lavoisier expressions) with standard fishbone diagrams. These diagrams then are used as input for the Papin specification.

\emph{Jet}~\cite{Ackermann:2012:JEDH} is an embedded DSL written in Scala which is meant to be used for batch processing of large datasets. This DSL can generate code for both Apache Spark and Apache Hadoop, both of which are systems for processing big data. The compiler, relational, and domain-specific optimisations of the implementation provide substantial performance gains compared to the naive handwritten processing code. Compared to \CPSLint, Jet is much more focused on high performance.
 
\emph{DSL4DPiFS}~\cite{Vogel-Heuser:2025:GNMD} is a graphical DSL to help with deploying data pipelines in industrial metal forming systems, built with interdisciplinary collaboration in mind. These data-pipelines are used to enable process optimisation, quality management, and predictive maintenance of these systems. For \CPSLint, we consider the data-pipeline to be an external factor where \CPSLint-produced code gets integrated.

\emph{KNIME}~\cite{Berthold:2009:KNIME} is a graphical platform for building data processing and analytics workflows from interconnected nodes. It supports a broad range of data integration, transformation, analysis, and visualisation tasks in a general-purpose workflow environment. KNIME offers substantially broader workflow support than \CPSLint but also requires a longer learning time.

\Cref{tab:related_work_comparison} shows the position of \CPSLint relative to prior work. The comparison indicates that the related tools and DSLs overlap with \CPSLint only partially. Some focus on querying, some on monitoring, some on documentation, optimisation, or deployment, but none combines declarative sanitisation, imputation, and compartmentalisation for industrial CPS time-series data in the same way, and only one covers a form of data imputation.
\begin{table*}[htbp]
    \caption{Comparison of supported functionality amongst existing tools/DSLs and with \CPSLint in the last row. The considered markings and their meanings are: \CIRCLE = supported, \LEFTcircle = unclear/partially supported, and \Circle = unsupported.}
    \label{tab:related_work_comparison}
    \begin{tabular}{@{}lcccccc@{}}
        \toprule 
        \textbf{Tool/DSL}       & \textbf{Imputation}   & \textbf{Type inference}   & \textbf{Filtering}    & \textbf{Statistical analysis}     & \textbf{Data restructuring} \\ 
        \midrule
        GNU datamash            &\Circle                &\Circle                    &\Circle                &\CIRCLE                            &\CIRCLE      \\ 
        Lisp Query Notation     &\Circle                &\Circle                    &\CIRCLE                &\Circle                            &\CIRCLE      \\ 
        DescribeML              &\Circle                &\CIRCLE                    &\Circle                &\Circle                            &\Circle      \\ 
        RADAR                   &\Circle                &\LEFTcircle                &\CIRCLE                &\CIRCLE                            &\LEFTcircle  \\ 
        Lavoisier               &\Circle                &\CIRCLE                    &\LEFTcircle            &\Circle                            &\CIRCLE      \\ 
        Jet                     &\Circle                &\Circle                    &\LEFTcircle            &\Circle                            &\CIRCLE      \\ 
        DSL4DPiFS               &\Circle                &\LEFTcircle                &\Circle                &\Circle                            &\LEFTcircle  \\ 
        KNIME                   &\LEFTcircle            &\LEFTcircle                &\CIRCLE                &\CIRCLE                            &\CIRCLE      \\
        \hline
        \CPSLint                &\CIRCLE                &\CIRCLE                    &\CIRCLE                &\Circle                            & \CIRCLE     \\ 
        \bottomrule
    \end{tabular}
\end{table*}

\section{Conclusion and future work}
\label{sec:conclusion}
We have elaborated the design and implementation of \CPSLint, a custom DSL written in the Rascal language workbench. Our DSL is intended to facilitate data sanitisation and compartmentalisation in industrial CPS use-cases, where workflows process data collections in time-series format. We have provided examples of such workflows, in which data parsing, data cutting and ad hoc data sanitisation stages can be collectively replaced by processing through \CPSLint. \CPSLint as a tool and in its integrated form with workflows from \Cref{sec:implementation}, can be considered as both a \emph{proof of concept} and a \emph{proof of integration} (based on the definitions by Odyurt et al.~\cite{Odyurt:2026:DICP}) for the ZORRO project.

In its current form, \CPSLint is a self-contained tool with sufficient functionality for our demonstration. It can also be easily extended and adapted to the needs of other use-cases, for instance by defining different data headers, or by implementing extra data imputation methods. Such additions can be considered as future extensions.

\begin{acknowledgments}
This publication is part of the project ZORRO\footnote{\url{https://zorro-project.nl}} with project number KICH1.ST02.21.003 of the research programme Key Enabling Technologies (KIC), which is (partly) financed by the Dutch Research Council (NWO).
\end{acknowledgments}

\begin{aideclaration}
 During the preparation of this work, the authors used ChatGPT-5.X in order to perform grammar and spelling checks. Further, the authors used ChatGPT Codex for setting up the CEURART template. After using these tools, the authors reviewed and edited the content as needed and take full responsibility for the publication’s content. 
\end{aideclaration}



\bibliography{bibliography/references}

@inproceedings{Odyurt:2021:PPFT,
    author = {Odyurt, Uraz and Roeder, Julius and Pimentel, Andy D. and Alonso, Ignacio Gonzalez and de Laat, Cees},
    title = {{Power Passports for Fault Tolerance: Anomaly Detection in Industrial CPS Using Electrical EFB}},
    booktitle = {2021 4th IEEE International Conference on Industrial Cyber-Physical Systems (ICPS)},
    year = {2021},
    OPTpages = {152--157},
    doi = {10.1109/ICPS49255.2021.9468262}
}

@article{Mernik:2005:WHDD,
    author = {Mernik, Marjan and Heering, Jan and Sloane, Anthony M.},
    title = {{When and How to Develop Domain-Specific Languages}},
    year = {2005},
    journal = {ACM Computing Surveys},
    OPTmonth = dec,
    OPTpages = {316--344},
    OPTnumpages = {29},
    OPTissue_date = {December 2005},
    OPTpublisher = {Association for Computing Machinery},
    OPTaddress = {New York, NY, USA},
    OPTvolume = {37},
    OPTnumber = {4},
    OPTissn = {0360-0300},
    doi = {10.1145/1118890.1118892}
}

@online{GNU:2025:Datamash,
    author = {{GNU Project}},
    title = {{GNU datamash}},
    year = {2025},
    url = {https://www.gnu.org/software/datamash/},
    urldate = {2025-08-26}
}

@misc{Ackermann:2012:JEDH,
    author = {Ackermann, Stefan and Jovanovic, Vojin and Rompf, Tiark and Odersky, Martin},
    title = {{Jet: An Embedded DSL for High Performance Big Data Processing}},
    year = {2012}, 
    url = {https://infoscience.epfl.ch/handle/20.500.14299/85985}
}

@inproceedings{Heine:2020:DADQ,
    author = {Heine, Felix and Kleiner, Carsten and Oelsner, Thomas},
    OPTeditor = {Hartmann, Sven and K{\"u}ng, Josef and Kotsis, Gabriele and Tjoa, A. Min and Khalil, Ismail},
    title = {{A DSL for Automated Data Quality Monitoring}},
    booktitle = {Database and Expert Systems Applications},
    year = {2020},
    OPTpublisher = {Springer International Publishing},
    OPTaddress = {Cham},
    OPTpages = {89--105},
    OPTisbn = {978-3-030-59003-1},
    doi = {10.1007/978-3-030-59003-1_6}
}

@article{Vogel-Heuser:2025:GNMD,
    author = {Birgit Vogel-Heuser and Mingxi Zhang and Marius Krüger and Alejandra Vicaria and Markus Gardill and Yuyao Jiang and Ansgar Trächtler and Henning Peters and Mathias Liewald and Adrian Schenek and Pascal Heinzelmann and Michael Weyrich},
    title = {{DSL4DPiFS --- A Graphical Notation to Model Data Pipeline Deployment in Forming Systems}},
    journal = {at - Automatisierungstechnik},
    year = {2025},
    OPTpages = {232--250},
    OPTvolume = {73},
    OPTnumber = {4},
    doi = {10.1515/auto-2024-0114}
}

@misc{Hoff:2024:LQND,
    author = {Hoff, Anders},
    title = {{Lisp Query Notation — A DSL for Data Processing}},
    year = 2024,
    OPTmonth = may,
    OPTpublisher = {Zenodo},
    doi = {10.5281/zenodo.11001584}
}

@article{Giner-Miguelez:2023:DSLD,
    author = {Joan Giner-Miguelez and Abel Gómez and Jordi Cabot},
    title = {{A Domain-Specific Language for Describing Machine Learning Datasets}},
    journal = {Journal of Computer Languages},
    year = {2023},
    OPTvolume = {76},
    OPTpages = {101209},
    OPTissn = {2590-1184},
    doi = {10.1016/j.cola.2023.101209}
}

@article{Sal:2024:DSLA,
    author = {Brian Sal and Diego García-Saiz and Alfonso {de la Vega} and Pablo Sánchez},
    title = {{Domain-Specific Languages for the Automated Generation of Datasets for Industry 4.0 Applications}},
    journal = {Journal of Industrial Information Integration},
    year = {2024},
    OPTvolume = {41},
    OPTpages = {100657},
    OPTissn = {2452-414X},
    doi = {10.1016/j.jii.2024.100657}
}

@article{DeLaVega:2020:LDIL,
    author = {Alfonso {de la Vega} and Diego García-Saiz and Marta Zorrilla and Pablo Sánchez},
    title = {{Lavoisier: A DSL for Increasing the Level of Abstraction of Data Selection and Formatting in Data Mining}},
    journal = {Journal of Computer Languages},
    year = {2020},
    OPTvolume = {60},
    OPTpages = {100987},
    OPTissn = {2590-1184},
    doi = {10.1016/j.cola.2020.100987}
}

@inproceedings{Klint:2009:RDSL,
    author = {Klint, Paul and van der Storm, Tijs and Vinju, Jurgen},
    booktitle = {2009 Ninth IEEE International Working Conference on Source Code Analysis and Manipulation}, 
    title = {{RASCAL: A Domain Specific Language for Source Code Analysis and Manipulation}}, 
    year = {2009},
    OPTpages = {168--177},
    doi = {10.1109/SCAM.2009.28}
}

@inproceedings{Klint:2011:EMPR,
    author = {Klint, Paul and van der Storm, Tijs and Vinju, Jurgen},
    OPTeditor = {Fernandes, Jo{\~a}o M. and L{\"a}mmel, Ralf and Visser, Joost and Saraiva, Jo{\~a}o},
    title = {{EASY Meta-programming with Rascal}},
    booktitle = {Generative and Transformational Techniques in Software Engineering III: International Summer School, GTTSE 2009, Braga, Portugal, July 6-11, 2009. Revised Papers},
    year = {2011},
    publisher = {Springer},
    OPTaddress = {Berlin, Heidelberg},
    OPTpages = {222--289},
    OPTisbn = {978-3-642-18023-1},
    doi = {10.1007/978-3-642-18023-1_6}
}

@software{Sayilir:2025:CODE,
	author = {Odyurt, Uraz and Sayilir, \"{O}mer and Stoelinga, Mariëlle and Zaytsev, Vadim},
	title = {{CPSLint}},
	month = apr,
	year = 2026,
	publisher = {Zenodo},
	version = {v1.0.0},
	doi = {10.5281/zenodo.17406795},
	OPTurl = {https://doi.org/10.5281/zenodo.17406795}
}

@inproceedings{Odyurt:2021:CAMA,
    author = {Odyurt, Uraz and Sapra, Dolly and Pimentel, Andy D.},
    OPTeditor = {Fujita, Hamido and Selamat, Ali and Lin, Jerry Chun-Wei and Ali, Moonis},
    title = {{The Choice of AI Matters: Alternative Machine Learning Approaches for CPS Anomalies}},
    booktitle = {Advances and Trends in Artificial Intelligence. From Theory to Practice},
    year = {2021},
    OPTpublisher = {Springer International Publishing},
    OPTaddress = {Cham},
    OPTpages = {474--484},
    OPTisbn = {978-3-030-79463-7},
    doi = {10.1007/978-3-030-79463-7_40}
}

@inproceedings{Odyurt:2026:DSLP,
    author = {Uraz Odyurt and Ömer Sayilir and Mariëlle Stoelinga and Vadim Zaytsev},
    title = {{CPSLint: A Domain-Specific Language Providing Data Validation and Sanitisation for Industrial Cyber-Physical Systems}}, 
    booktitle = {Proceedings of the 19th ACM SIGPLAN International Conference on Software Language Engineering (SLE)},
    year = {2026},
	publisher = {ACM},
    doi = {10.1145/3806383.3815519}
}

@misc{Odyurt:2026:DICP,
    author = {Uraz Odyurt and Richard Loendersloot and Tiedo Tinga},
    title = {{Demonstrators for Industrial Cyber-Physical System Research: A Requirements Hierarchy Driven by Software-Intensive Design}},
    year = {2026},
    OPTeprint = {2510.18534},
    OPTarchivePrefix = {arXiv},
    OPTprimaryClass = {cs.SE},
    doi = {10.48550/arXiv.2510.18534}
}

@article{Erdweg:2015:ECLW,
    author = {Sebastian Erdweg and Tijs van der Storm and Markus V{\"{o}}lter and Laurence Tratt and Remi Bosman and William R. Cook and Albert Gerritsen and Angelo Hulshout and Steven Kelly and Alex Loh and Gabri{\"{e}}l D. P. Konat and Pedro J. Molina and Martin Palatnik and Risto Pohjonen and Eugen Schindler and Klemens Schindler and Riccardo Solmi and Vlad A. Vergu and Eelco Visser and Kevin van der Vlist and Guido Wachsmuth and Jimi van der Woning},
    title = {{Evaluating and comparing language workbenches: Existing results and benchmarks for the future}},
    journal = {Computer Languages, Systems and Structures},
    year = {2015},
    OPTvolume = {44},
    OPTpages = {24--47},
    doi = {10.1016/J.CL.2015.08.007}
}

@article{Berthold:2009:KNIME,
    author = {Berthold, Michael R. and Cebron, Nicolas and Dill, Fabian and Gabriel, Thomas R. and K\"{o}tter, Tobias and Meinl, Thorsten and Ohl, Peter and Thiel, Kilian and Wiswedel, Bernd},
    title = {{KNIME --- the Konstanz information miner: version 2.0 and beyond}},
    journal = {SIGKDD Explorations Newsletter},
    year = {2009},
    OPTissue_date = {June 2009},
    OPTpublisher = {ACM},
    OPTaddress = {New York, NY, USA},
    OPTvolume = {11},
    OPTnumber = {1},
    OPTissn = {1931-0145},
    OPTmonth = {nov},
    OPTpages = {26--31},
    OPTnumpages = {6},
    doi = {10.1145/1656274.1656280}
}

\end{document}